\documentclass[a4paper,10pt]{article}
\usepackage[dvipsnames]{xcolor}
\usepackage[utf8]{inputenc}

\title{  Dualized  Gravity beyond Linear Approximation }
\author{  Salman Sajad Wani$^1$,  Tsou Sheung Tsun$^{2}$, Mir Faizal$^3$  \\ \\
$^{1, 3}$Canadian Quantum Research Center (CQRC)\\
204-3002 32 Ave Vernon, BC, V1T 2L7 Canada\\
$^{1}$School of Physics, Damghan University,\\ P.O. Box 3671641167, Damghan, Iran
\\
$^2$Mathematical Institute, University of Oxford,\\ Andrew Wiles
Building, \\Radcliffe Observatory Quarter, Woodstock Road,
\\ Oxford
OX2 6GG, United Kingdom 
\\
$^3$Irving K. Barber School of Arts and Sciences, \\ University of British Columbia - Okanagan,
\\ Kelowna, British Columbia V1V 1V7, Canada
\\$^3$Department of Physics and Astronomy,\\
University of Lethbridge, \\
Lethbridge, Alberta, T1K 3M4, Canada
 }
\date{}

\begin{document}

\maketitle

\begin{abstract}
We will construct a loop space formalism for general relativity, and 
construct the  Polyakov variables as connections for such a loop space. 
We will use these Polyakov variables to construct a dual theory of gravity 
beyond linear approximation. It will be demonstrated that this loop space 
duality reduces to the Hodge duality for linearized gravity.  Furthermore, 
a  loop space curvature will be constructed  from this Polyakov variable. 
It will be  shown that this loop space curvature vanishes in the
absence of topological defects, and so it can be used to investigate 
gravitational monopoles.  We will also construct the suitable monopole 
charge for such gravitational monopoles.  
\end{abstract}
\section{Introduction}
The electromagnetic  Hodge duality has been used to   obtain  interesting results  \cite{emhd, emhd1}, and so it is interesting to investigate such a Hodge duality for gravity. As  the linearized gravity can be analyzed using gravitoelectromagnetism \cite{ge01, ge02, ge12, ge14},   the Hodge duality has been constructed for  linearized gravity, and it has also  been used to  obtain  a dual gravitational theory \cite{du12, du14, du15, du16}. 
This Hodge duality for linearized gravity has been generalized to curved spacetime \cite{cu12, cu14}.  The dual gravity has been used in M-theory \cite{mt12, mt14}. It has been proposed that the M-theory could have a $(4, 0)$ phase in six dimensions \cite{mt16}.
The free $(4,0)$ six dimensional  theory is obtained  from  five dimensional  linearized supergravity theory. 
The  reduction of this theory to four dimensions on a $2$-torus has an
$SL(2, Z)$ duality symmetry, which   acts   on the linearized gravitational  sector of the theory.  This 
interchanges the  Bianchi identities and the linearized Einstein equations \cite{mt12, mt14}. Thus, it produces    a self-duality between strong and weak coupling regimes of the theory. It may be noted that the Hodge dual of linearized gravity has  been studied in $E11$ generalized eleven dimensional  geometry \cite{mt01}. 
It has also been used to investigate $E7$ generalized  eleven dimensions 
geometry \cite{mt02}. Thus, important results in M-theory have been obtained 
using the gravitational version of Hodge duality for linearized gravity. 
It would be interesting to generalize this duality  to full gravitational 
theory beyond its linearized limit. However, as it would contain non-linear 
terms, and it would resemble Yang-Mills theory. In fact, it is known that  
in the vierbein formalism,  general relativity can be viewed as a gauge theory 
of Lorentz group, with spin connection  as its gauge  connection \cite{dynam2, dynam0,  dynam4, dynam6,  dynam, dynam1}. It may be noted that it has been proposed that we can solve problems with perturbative quantum gravity by using gauge connection as the dynamical variable \cite{dynam}.  The dimension reduction has been investigated using a dynamical gravitational Higgs mechanism with spin connection as the dynamical variable \cite{dynam1}. Thus, we can use the spin connection to construct such a dual model for gravity   beyond its linearized approximation. 

The problem with this approach  is that it has not been possible to 
construct a generalization of Hodge duality to non-abelian gauge theories 
in spacetime. However, it is possible to construct a such a duality    
in  Yang-Mills theories using Polyakov variable  \cite{pq1,dual, dual1,  pq2}.
It has also been demonstrated that this  duality reduces to the        
electromagnetic Hodge duality for abelian gauge theories \cite{pq1}. 
These Polyakov variables are constructed using   Polyakov loops in loop  
space,  which are  holonomies of closed loops in spacetime \cite{p1, t1, t2,Chan:1995xr,  p, p1}. 
 These  Polyakov loops in loop  space  resemble the Wilson loops, as they are   parameterization independent.  However, unlike the 
Wilson loop, a trace is not  taken over the gauge group for the  Polyakov 
loops in the loop space.  So, these  Polyakov loops  are  gauge group-valued 
functions rather than simple a numbers   \cite{p1}.  The  connection in the 
Polyakov loop space is called Polyakov variable.
As  this Polyakov variable which has been used to construct a non-abelian generalization of Hodge duality \cite{pq1,dual, dual1,  pq2}, and gravity can be constructed as a gauge theory of Lorentz group 
\cite{dynam2, dynam0,  dynam4, dynam6,  dynam, dynam1}, we can construct Polyakov variable for gravity, and use it to construct a dual for gravity beyond linearized approximation.  
It may be noted that this duality has been used in  Yang-Mills theories to 
obtain several interesting results. These results have been obtained using  
a  non-abelian  dual potential for Yang-Mills theories, which has been 
constructed using this duality \cite{pq4, p9}. In fact, this    
non-abelian dual potential has been used for  construction of a  
Dualized Standard Model   \cite{pq4, p9, pq0, q01, d0}. This  Dualized 
Standard Model has been used for investigating   the mass difference  
between   various  generations of  fermions  \cite{d1, d2}.
 The  Neutrino oscillations have also been studied using this dual 
potential \cite{no}. 
  The Dualized Standard Model has been used to  investigate the    
off-diagonal elements of the CKM matrix  \cite{cmk}.
 As the construction of the  't Hooft's  order-disorder parameters require a non-abelian  dual potential, the Polyakov loop space formalism can be used to construct 't Hooft's  order-disorder parameters \cite{1t, d, 1p}.

The loop space curvature can be constructed using this Polyakov variable. This loop space curvature vanishes in absence of a monopole. Thus, it can be used to detect the existence of monopoles in spacetime. It has been possible to define a loop in this space of loops, and use it to obtain the charge for non-abelian monopoles \cite{p1, t1}. 
It may be noted that such monopoles have been studied for various theories using loop space formalism. 
The superspace formalism has been used to supersymmetrize the  Polyakov loop space,  and this  suersymmetric  loop space  has been used to analyze monopoles in supersymmetric gauge theories \cite{pqaa, pq1a}. This is because it has been  demonstrated that the supersymmetric  loop space curvature would also  vanish in absence of monopoles. Thus, it can be used to analyze topological defects in 
supersymmetric gauge theories. This supersymmetric loop space has also been 
used to construct a dual potential in supersymmetric gauge theories \cite{dual}. This has been  generalized to a  deformed supersymmetric gauge theory using deformed superspace \cite{dual1}.
Polyakov loops space has also been used for investigating the  
topological defects in  fractional M2-branes \cite{m2frac}.  
It has been demonstrated that the Polyakov loops for M2-branes can be 
constructed using a supergauge potential, with a supergroup as its gauge group. 
The Polyakov loop space for M2-branes has been used to analyze 
topological defects of M2-branes. The topological defect has also been 
studied in  a deformed gauge theory using Polyakov  loops \cite{td12}. 
This is done  by first deforming the gauge theory by minimal length in the 
background spacetime. This deformation of the gauge theory also deforms 
the Polyakov variables, which in turn deforms the loop space curvature. 
Thus, deformed Bianchi identity for such  deformed  gauge theories can be 
violated even if the original Bianchi identity is not violated. 
This can be viewed as the formation  of topological defects from minimal 
length in the background geometry of spacetime. Thus, the geometry of 
spacetime can have important consequences for topological defects. So, it would be important to analyze Polyakov loops in general relativity. 

 It is also important to investigate Polyakov loop space formalism for general relativity as it can be used to  analyze topological defects 
 in general relativity. It may be noted that it is possible to construct  
gravitational monopole solutions in analogy with the usual monopole solutions. 
In fact, it has been demonstrated that  dual supertranslation gauge symmetry can be used to construct a gravitational  Wu-Yang monopole solution \cite{gm12}. This is done using a specific metric describing the monopole solution, which is constructed from  two overlapping patches on a sphere. This  solution is separately regular on  the two  patches  and differentiable in the region where they are  overlapped. The gravitational   monopole solutions have also been studied in  K-theory \cite{gm14}. This is done by   reconstructing  the moduli space  for such solutions using  the NHD construction. 
The properties of   the Higgs field for a gravitational global monopoles have been investigated, and  a  bound for the maximum   vacuum value of such a  Higgs field has been  obtained \cite{gm15}.  
The   spherical symmetry solutions for  for gravitating global monopoles have
   been constructed, and used to investigate models of topological 
inflation \cite{gm16}. The gravitational monopoles have also been studied 
using the   twistors for  the moduli space of solutions. \cite{gm17}. 
These results have been obtained by analyzing the gravitational analogue
to Yang-Mills  monopoles. However, as Polyakov loops can be used to properly 
analyze non-abelian monopoles  \cite{p1, t1}, and general relativity can be viewed as a gauge theory of Lorentz group \cite{dynam2, dynam0,  dynam4, dynam6,  dynam, dynam1},  it   is important to construct gravitational Polyakov loop space. 

\section{Polyakov Variable}
The general relativity can be viewed as 
  a gauge theory of the Lorentz group, with  the
spin connection as a gauge potential \cite{dynam2, dynam0,  dynam4, dynam6,  dynam,
 dynam1}. This can be done 
by writing the metric in terms of the vierbein
 as $g_{\mu\nu} = e_{\mu b} e_{a\nu} \eta^{ab}$, and so $e=\sqrt{|g|}$. Now the spin connection, which can act as gauge potential can be written as   $ \omega_\mu^{ab}= -e^{\nu a}\nabla_\mu e^b_\nu = - e^{\nu a}\partial_\mu e^b_\nu +  e^{\nu a}\Gamma_{\mu\nu}^\rho e^b_\rho$, with $\Gamma_{\mu\nu}^\rho$ as the  Levi-Civita connection. We can also use  the generators of the Lorentz group $\Sigma_{ab} = - i 
[\gamma_a, \gamma_b]/4$, with $\gamma_a, \gamma_b$ as Dirac matrices, 
and write $\omega^\mu = \omega^{\mu ab} \Sigma_{ab} $ as the gauge potential.  So, using $\omega^\mu$ as the gauge potential, we  can define the gauge covariant derivative as $D_\mu = \partial_\mu - i \omega_\mu$. Now under Lorentz 
transformation $e_\mu^a \to \Lambda^a_b e^b_\mu$, the spin connection  $\omega_\mu$ transforms as $\omega_\mu \to U \omega_\mu 
U^{-1} - (\partial_\mu U) U^{-1}$, with $ U = \exp i \Lambda^{ab}  \Sigma_{ab}/2$. 

Now if we can define a field strength as  $F =\Sigma_{ab} F^{ab}_{\mu\nu}  dx^\mu \wedge dx^\nu $ and $\omega = \omega_\mu dx^\mu$, with $F= D\wedge \omega = d \omega + \omega \wedge \omega$. So,  it is possible to express  the field strength of the spin connection as \cite{dynam2, dynam0,  dynam4, dynam6,  dynam, dynam1}
\begin{equation}
 F_{\mu\nu}^{ab} = \partial_\mu \omega_\nu^{ab}
 -  \partial_\nu \omega_\mu^{ab} + \omega_\mu^{ac} \omega^b_{\nu c}- \omega_\nu^{ac} \omega^b_{\mu c}. 
\end{equation}
This the field strength transforms as $F^{ab}_{\mu\nu} \to U F^{ab}_{\mu\nu}  U^{-1} $. 
It may be noted that this  field strength is the  Riemannian curvature tensor $ R_{\mu\nu}^{ab}  \equiv F_{\mu\nu}^{ab} 
$,  because we can write the  Riemannian curvature tensor in metric formalism as $  g^{\rho \sigma} R^{\tau}_{\rho \mu\nu} =  e^\tau_a e^\sigma_b   F_{\mu\nu}^{ab} $.  So, we can also write the scalar curvature as $R = e_a^\mu e_b^\nu F^{ab}_{\mu\nu}$, and use it to construct the Einstein Hilbert action 
\begin{equation}
S = \int d^4 x e R. 
\end{equation}
It has been suggested that the spin connection can be used  as a dynamical gauge connection in perturbative quantum gravity \cite{dynam}. In fact, even the  Higgs mechanism has been studied with  spin connection as the dynamical variable \cite{dynam1}. So, here we will construct the loop space using spin connection as the gauge connection. 

This can be done by  considering the space of loops in  spacetime, with a fixed base point.
Now a  loop is   parameterized  by the coordinates $\xi^\mu(s)$, 
\begin{equation}
 C : \{ \xi^\mu (s): s = 0 \to 2\pi, \, \, \xi^\mu (0) = \xi^\mu(2\pi)\},  
\end{equation}
where    $\xi^\mu (0) = \xi^\mu(2\pi)$ is the chosen (but arbitrary) base point \cite{p1, t1,Chan:1995xr,  t2, p}.  Next we define the loop space variable
\begin{equation}
   \Phi     [\xi]  = 
P_s \exp i \int^{2\pi}_0  \omega^\mu (\xi(s)) \frac{d \xi_\mu}{ds}. 
\end{equation}
where $P_s$ denotes ordering in $s$,
increasing from right to left.  So,  we can define its
logarithmic functional derivative as  a   loop space connection
\begin{equation}
 F_\mu [\xi| s] = i   \Phi    ^{-1}[\xi] \frac{\delta}{\delta \xi^\mu (s)}  \Phi    [\xi]. 
 \end{equation}
The derivative in $s$ is  taken from below. It may be noted as the loop variable $  \Phi     [\xi]$
only depends on $C$ and not the manner in which $C$ is parameterized, so labeling it with a fixed point is over 
complete.  In fact, any other parameterization of $C$ will  
only change the variable in the integration and not the loop space variable $  \Phi     [\xi]$. 

We can obtain a expression  relating $ F_\mu [\xi| s]$ to the spacetime curvature $R^{\mu\nu ab} \Sigma_{ab}$ by first defining a parallel transport 
 from a point $\xi(s_1)$ to a point $\xi (s_2)$ as   
\begin{eqnarray}
   \Phi     [\xi: s_1, s_2 ]  =  
P_s \exp i \int^{s_2}_{s_1} \omega^\mu (\xi(s)) \frac{d \xi_\mu}{ds}. 
\label{transport}
\end{eqnarray}
Thus, we can write the loop space connection as 
\begin{equation}
F^\mu[\xi|s]=  \Phi    ^{-1} [\xi: s,0] R^{\mu\nu ab} \Sigma_{ab}   \Phi     [\xi: s,0]
\frac{d\xi_\nu(s)}{ds}.
\label{looptolocal}
\end{equation}
So, we  parallel transport  from a fixed point  along a fixed path  to another fixed point. After reaching that point, 
we took  a  detour then turned
back along the same path till we reach the  original point. 
The phase factor generated by 
going along the original path   will be canceled by the phase factor generated by coming  back along it.   However, there will be a  contribution generated by  the transport along the infinitesimal circuit  along the  final point. This contribution is proportional to the spacetime curvature at that point.  This is similar to the procedure of defining the loop space curvature in Yang-Mills theories \cite{p1, t1, Chan:1995xr, t2, p}. 
 
\section{Topological Defects }
The topological defects can be investigated using  loop space formalism. We first observe that the    $ {F}_\mu  [\xi|s] $ represents the     connection in this loop space.  Now   this connection  $ {F}_\mu  [\xi|s] $,  can be used to define   a functional covariant  derivative as    $\Delta_\mu (s) =  \delta/\delta \xi^\mu (s) + i  {F}_\mu  [\xi|s] 
$. This functional covariant derivative can be used to construct a loop space curvature term $ {G}_{\mu\nu}[\xi, s_1, s_2]$. It may be noted that the commutator of two such functional covariant  derivatives can be expressed as $ -i {G}_{\mu\nu}[\xi, s_1, s_2]$. So, we can write the loop space curvature as     
\begin{eqnarray}
{i [\Delta_\mu (s_1), \Delta_\nu (s_2)] =}  {G}_{\mu\nu}[\xi ( s_1, s_2)] &=&  \frac{\delta}{\delta \xi^\mu (s_2) } {F}_\nu  [\xi|s_1]
- \frac{\delta}{\delta \xi^\nu (s_1) } {F}_\mu  [\xi|s_2] \nonumber \\&&
+i [ {F}_\mu  [\xi|s_1], {F}_\nu  [\xi|s_2]], 
\end{eqnarray}

Now topological defects  occurring due to gravitational monopoles can be 
investigated using this    loop space curvature. The gravitational Bianchi 
identity gets violated, if a gravitational monopole is present in the system. 
So, the violation of the gravitational Bianchi identity can be taken as an 
indication for the presence of gravitational monopoles. This violation can in 
turn be related to the loop space curvature. This can be done as the       
oop space curvature is formed from an   infinitesimal  circuit. So,   we can  
start from  one point in loop space  move  in a certain direction to another 
point in loop space.  After reaching that point,  we can travel in a different 
direction. We return by first moving in  the first direction, followed by the 
second direction. Now by this motion in loop space,  we can obtain the loop space 
curvature. Thus, to obtain the loop space curvature, we   consider variations 
of the curve in two directions $\lambda$ and $\kappa$. 
Specifically,
consider the three curves \cite{t1}:  
\begin{eqnarray}
\xi_1^\mu (s)&=& \xi^\mu (s) + \delta_1 (\xi^\mu (s))
                   \nonumber\\
\xi_2^\mu (s) &=& \xi^\mu (s) + \delta_2 (\xi^\mu (s))
\nonumber\\
\xi_3^\mu (s) &=& \xi_1^\mu (s) + \delta_2(\xi^\mu(s)),
\end{eqnarray}
with the variations $\delta_1$ at $s_1$ in the direction $\lambda$:
\begin{equation}
\delta_1 (\xi^\mu (s)) =\Delta \delta^\mu_\lambda \delta(s-s_1),
\label{delta1}
\end{equation}
and the variation $\delta_2$ at $s_2$ in the direction $\kappa$:
\begin{equation}
\delta_2 (\xi^\mu (s)) = \Delta' \delta^\mu_\kappa \delta(s-s_2),
\label{delta2}
\end{equation}
taking first the limit $\Delta' \to 0$, and then the limit $\Delta \to 0$. 
 
With respect to these variations, we can write down the
the functional  derivative of ${F}_\lambda [\xi|s_1]$ 
\begin{equation}
\frac{\delta}{\delta \xi^\kappa(s_2)}  {F}_\lambda [\xi|s_1] = \lim_{\Delta
  \to 0} \lim_{\Delta' \to 0} \frac{1}{\Delta \Delta'} \frac{i}{g}
\left\{ \Phi ^{-1}[\xi_2] \Phi [\xi_3] -
  \Phi ^{-1}[\xi] \Phi [\xi_1] \right\},
\end{equation}
where the indices $\kappa, \lambda$, and the paramers $s_1, s_2$ on the left have been incorporated on the right through the definitions (\ref{delta1}) and (\ref{delta2}).

We need to obtain the value of   $    \Phi       ^{-1}[\xi_2]    \Phi       [\xi_3] 
-    \Phi       ^{-1}[\xi]   \Phi     [\xi_1]$ to  find an explicit expression for the loop space curvature. We can use  parallel
transport along these paths in loop space to obtain 
\begin{equation}
    \Phi     [\xi_1] =   \Phi     [\xi] - i \int ds   \Phi     [ \xi: 2\pi, s ]
   {F} (\xi(s))    \Phi     (\xi: s, 0). 
\end{equation}  
Here    ${F} (\xi(s))$ can be written as  $
   {F}  (\xi(s))  =
   R^{\mu\nu ab} \Sigma_{ab}(\xi(s)) ({d \xi_{ \nu  }(s)}/{ds}) 
\Delta  \delta_\mu^\lambda \delta(s-s_1).  
$
It is also possible to write $   \Phi     [\xi_2]$ as  
\begin{equation}
    \Phi     [\xi_2] =   \Phi     [\xi] - i \int ds   \Phi     [ \xi: 2\pi, s ]
   {F}  (\xi(s))    \Phi     [\xi: s, 0], 
\end{equation}
Here $   {F}  (\xi(s))$ can be expressed as  $   {F}  (\xi(s))  =
     R^{\mu\nu ab} \Sigma_{ab}(\xi(s)) ({d \xi_{ \nu  }/(s)}){ds} 
\Delta' \delta_\mu^\kappa \delta (s-s_2)$.   
 Next we consider $   \Phi     [\xi_3]$, and write   
\begin{equation}
    \Phi     [\xi_3] =    \Phi     [\xi_1] - i \int ds   \Phi     [ \xi_1: 2\pi, s ]
   {F} (\xi_1(s))   \Phi     [\xi_1: s, 0].
\end{equation}
Here we can write  ${F} (\xi_1(s))$  as $   {F}  (\xi_1(s))  =
   R^{\mu\nu ab} \Sigma_{ab} (\xi_1(s)) ({d \xi_{1 \nu  }/(s)}){ds} 
\Delta' \delta^\kappa_\mu \delta (s-s_2).$ 
We can also obtain the value of   $  \Phi     [\xi: 2\pi, s]$ and 
$  \Phi     [\xi_1: s, 0]$. So, using all these expression, we can write the functional derivatives of the loop space connection as 
\begin{eqnarray}
 \frac{\delta }{\delta \xi_\mu (s_2)}  {F}_\nu  [\xi | s_1] &=& 
   \Phi     ^{-1}[\xi: s_1, 0]    {D}^\nu R^{\mu\rho ab}\Sigma_{ab} (\xi
 (s_2)) \nonumber \\ 
&& \times\frac{d\xi_\rho (s_1)}{d s_1}  \Phi      [\xi: s_1, 0] \delta (s_2 - s_1)\nonumber \\ && + 
   \Phi     ^{-1}[\xi: s_2, 0]   R_{\mu\nu}^{ab}\Sigma_{ab}(\xi (s_2)) 
  \Phi      [\xi: s_2, 0]\nonumber \\ && \times \frac{d}{d s_1}\delta (s_2 - s_1)
 \nonumber \\ && +i [ {F}_\mu  [\xi|s_2],  {F}_\nu  [\xi|s_1]]\theta (s_1 - s_2). 
\nonumber \\
 \frac{\delta }{\delta \xi_\nu (s_1)}  {F}_\mu  [\xi | s_2] &=& 
   \Phi     ^{-1}[\xi: s_2, 0]    {D}^\mu  R^{\mu\tau ab}\Sigma_{ab}(\xi (s_1)) \nonumber \\ && \times \frac{d\xi_\tau (s_2)}{d s_2}  \Phi      [\xi: s_2, 0] \delta (s_1 - s_2)
 \nonumber \\ && + 
   \Phi     ^{-1}[\xi: s_1, 0]   R_{\nu\mu}^{ab}\Sigma_{ab}(\xi (s_1))   \Phi      [\xi: s_1, 0]\nonumber \\ && \times \frac{d}{d s_2}\delta (s_1 - s_2)
 \nonumber \\ && +i [ {F}_\nu  [\xi|s_1],  {F}_\mu  [\xi|s_2]]\theta (s_2 - s_1). 
\end{eqnarray}
Now as the loop space  curvature  was expressed in terms of  functional derivatives of loop space connections, we can express it as  
\begin{eqnarray}
 {G}_{\mu\nu}[\xi ( s_1, s_2)] &=& \frac{\delta}{\delta \xi^\mu (s_2) } {F}_\nu  [\xi|s_1]
- \frac{\delta}{\delta \xi^\nu (s_1) } {F}_\mu  [\xi|s_2] \nonumber \\&&
+i [ {F}_\mu  [\xi|s_1],  {F}_\nu  [\xi|s_2]]\nonumber \\ &=& 
   \Phi     ^{-1}[\xi: s_1,0] \Big[[   {D}_\mu ,                                   {R}_{\nu\tau}^{ab}\Sigma_{ab}]  +[   {D}_\nu ,  {R}_{\tau\mu}^{ab}\Sigma_{ab}] 
 + [   {D}_\tau ,  {R}_{\mu\nu}^{ab}\Sigma_{ab}] \Big]
 \nonumber \\ && \times 
   \Phi     [\xi: s_1,0]\frac{d\xi^\tau (s_1)}{ds} \delta (s_1-s_2). 
\end{eqnarray}
Thus, the loop space  curvature  is proportional  to the  gravitational  Bianchi identity. The presence  of monopoles will violate this gravitational  Bianchi identity, $[{D}_\mu ,  R_{\nu\tau}^{ab}\Sigma_{ab}]  +[   {D}_\nu ,  R_{\tau\mu}^{ab}\Sigma_{ab}] 
 + [   {D}_\tau ,  R_{\mu\nu}^{ab}\Sigma_{ab}]\neq0$. Thus, the loop space curvature will not vanish, when there is a gravitational monopole in spacetime.  However, in absence of such a gravitational monopole, the gravitational Bianchi identity will not be violated,   
$[   {D}_\mu ,  R_{\nu\tau}^{ab}\Sigma_{ab}]  +[   {D}_\nu ,  R_{\tau\mu}^{ab}\Sigma_{ab}] 
 + [   {D}_\tau ,  R_{\mu\nu}^{ab}\Sigma_{ab}] =0$. The loop space curvature will vanish, in absence of a gravitational monopole. So,  the loop space curvature can be used to  study gravitational monopoles.

\section{Generalized Dual Transform}
We will use the loop space variable to construct duality for general relativity. To analyze this duality, we define a new loop space variable   $E_\mu[\xi|s]$ 
as 
\begin{equation}
E_\mu[\xi|s] \textcolor{red}{}=   \Phi      [\xi: s,0] F_\mu[\xi|s]   \Phi    ^{-1}  [\xi: s,0]
\label{Emuxis}
\end{equation}
It may be noted here  the variable  $E_\mu[\xi|s]$ only 
  dependent on the segment of the loop $\xi$ between two points 
$s - \epsilon/2$ to $s + \epsilon/2$ (here the $\delta$ function has been replaced by a function with a finite width $\epsilon$.  Thus, we can now define  loop derivative for $E_\mu[\xi|s]$ using the standard procedure, and take the  limit $\epsilon \rightarrow 0$ at the end of our calculations. To construct a second loop derivative at the same point 
  $s$, we again consider the $\delta$ function as a function with a 
 finite width $\epsilon'$), and define the   second derivative
 on this segment of such a  loop.  Then after taking these two loop derivatives, we first  by take the 
  limit $\epsilon' \rightarrow 0$,  and the we take   the limit
$\epsilon \rightarrow 0$.  
It is possible to construct a loop space formalism for general relativity using   $E_\mu[\xi|s]$. 

Now using  a suitable Heaviside $\theta$-function, we  can write the functional derivative of $E_\mu[\xi|s]$  as
\begin{eqnarray}
 \frac{\delta }{\delta \xi^\nu (s_1) }E_\mu[\xi|s] &=&   \Phi     [ \xi : s,0] [ \frac{\delta }{\delta \xi^\nu (s_1) }  F_\mu[\xi|s]
  \\ && \nonumber + i \theta(s-s_1) [F_\nu[\xi|s'], F_\mu[\xi|s]] ]   \Phi    ^{-1}[ \xi : s, 0]  
\end{eqnarray}
This loop space connection can be used to construct the 
 loop space curvature as   
\begin{eqnarray}
G_{\mu\nu}[\xi : s_1, s_2 ] =   \Phi    ^{-1} [\xi: s,0] [\frac{\delta}{\delta \xi^\nu(s_1)} E_\mu[\xi|s_2]
 \nonumber \\   - \frac{\delta}{\delta \xi^\mu(s_2)} E_\nu[\xi|s_1]]   \Phi     [ \xi : s,0] 
\label{GmunuinE}
\end{eqnarray}
Now  we observe that $G_{\mu\nu}[\xi : s_1, s_2 ]$ vanishes,   in the absence of a monopole. This is because in absence of  a monopole, the functional derivatives of $E_\mu[\xi|s_2]$ satisfy 
\begin{equation}
\frac{\delta}{\delta \xi^\nu(s_1)} E_\mu[\xi|s_2]
   - \frac{\delta}{\delta \xi^\mu(s_2)} E_\nu[\xi|s_1] = 0.
\label{GausslawE}
\end{equation}
So,   the loop space description of general relativity can be constructed using  $E_\mu[\xi|s]$ is equivalent to its loop space description constructed using  $F_\mu[\xi|s]$.   

Now we can express a dual potential to a spin connection of the general relativity, and thus construct a dual gravitational theory. This can be  done by introducing a set of dual set of variables
$\tilde{E}_\mu[\eta|t]$ which would be dual to the original loop 
variable $E_\mu[\xi|s]$. Here $\eta$ is 
another parametrized loop with parameter $t$, which is different from the original  parametrized loop $\xi$ with the parameter $s$.
We can express this dual set of variable    $\tilde{E}_\mu[\eta|t]$ as 
\begin{eqnarray}
\omega^{-1}(\eta(t)) \tilde{E}_\mu[\eta|t] \omega(\eta(t)) &=& -\frac{2}{\bar{N}}
   \epsilon_{\mu\nu\rho\sigma} \frac{d \eta^\nu}{d t} \int \delta\xi ds    
   E^\rho[\xi|s] \frac{d \xi ^\sigma}{d s } \left( \frac{d \xi^\tau}{ds} \frac{d \xi_\tau}{ds}\nonumber\right)^{-1} \\ && \times   \delta(\xi(s)-\eta(t)),  
\end{eqnarray}
where $\bar{N}$
a normalization constant, which can be written as 
\begin{equation}
\bar{N} = \int_0^{2\pi} ds \prod_{s' \neq s} d^4\xi(s').
\label{Nbar}
\end{equation}
Here $\omega(x)$ is a transformation  matrix which transforms the general relativity to its dual theory.  

We need to identify suitable quantities  in general relativity, which will 
appear as    monopoles in the dual theory, defined using  $\tilde{E}_\mu[\eta|t]$. Furthermore, we would also require that monopoles in the original theory to be equivalent to such  quantities in the dual theory. It is known that in Yang-Mills theories, the action of  gauge   covariant derivative on the  the curvature tensor satisfies both these properties \cite{t1}. So, the action of gauge covariant derivative on the field strength of the  Yang-Mills theory can be identified with such a quantity. As general relativity can also be viewed as a gauge theory of Lorentz symmetry, we can define such a quantity by the action of the covariant derivative on the curvature tensor of general relativity, $\mathcal{Y}_\mu = D^\nu{R} _{\mu\nu}^{ ab} \Sigma_{ab}  $. It may be noted here that the covariant derivative is again defined with  the  spin connection as the gauge potential.  Now using the Polyakov loop space formalism, we can be observed that this quantity can be related to the   non-vanishing functional divergence, $[\delta/\delta  \xi_\mu (s)]F_\mu[\xi|s]$. However, as the loop space can also be constructed using 
$E_\mu[\xi|s]$, we can also relate it to  non-vanishing functional divergence of  $E_\mu[\xi|s] $ as  
\begin{equation}
 \frac{\delta}{\delta \xi^\mu (s)}E_\mu[\xi|s] =   \Phi     [\xi: s,0] \frac{\delta}{\delta \xi^\mu (s)} F_\mu[\xi|s]
     \Phi    ^{-1} [\xi: s,0].
\label{divEinF}
\end{equation}
In presence of a gravitational monopole,  in  general relativity    the loop space curvature 
$G_{\mu\nu}[\xi : s_1, s_2]$, does not vanish. This loop space curvature can now be expressed in terms of a functional  curl of $ E_\mu[\xi|s] $ as 
\begin{eqnarray}
G_{\mu\nu}[\xi : s_1, s_2 ] =   \Phi    ^{-1} [\xi: s,0] \{\frac{\delta}{\delta \xi^\nu(s_1)} E_\mu[\xi|s_2]
 \nonumber \\   - \frac{\delta}{\delta \xi^\mu(s_2)} E_\nu[\xi|s_1]\}   \Phi     [ \xi : s,0] 
\label{GmunuinE}
\end{eqnarray}
We have observed that the curl of $ E_\mu[\xi|s] $ can be used to investigate the presence of a monopole in general relativity. So, 
 the functional   curl of $\tilde{E}_\mu[\eta|t] $ can also  be used to investigate  the presence of a gravitational monopole in the dual gravitational theory. 
To show that $\mathcal{Y}_{\mu} = D^\nu{R} _{\mu\nu}^{ ab} \Sigma_{ab}$ in general relativity is a monopole in the dual gravity, we have to demonstrate that the  non-vanishing functional divergence of $ E_\mu[\xi|s_2]$ will produce  to a non-vanishing functional curl of the dual variable $\tilde{E}_\mu[\eta|t]$. 

Now the functional  divergence of the dual loop space variable  $ \tilde{E}_\mu[\eta|t]$ can be written expressed using the transformation matrix $ \omega(\eta(t))$. Thus, using a suitable transformation, we can relate it to the functional divergence of the loop space variable in general relativity as 
\begin{eqnarray}
& & \epsilon^{\lambda\mu\alpha\beta} \delta_\lambda(t)
   \Big\{\omega^{-1}(\eta(t)) \tilde{E}_\mu[\eta|t] \omega(\eta(t))\Big\} \nonumber\\
    & = & -\frac{2}{\bar{N}} \epsilon^{\lambda\mu\alpha\beta}
   \epsilon_{\mu\nu\rho\sigma} \frac{d \eta^\nu}{d t}  \int \delta\xi ds \nonumber \\  &&
   \Big\{ \frac{\delta}{\delta \xi_\lambda (s)} E^\rho[\xi|s]\Big\} \frac{d \xi^\sigma}{d s } \left( \frac{d \xi^\alpha}{d s }  \frac{d \xi_\alpha}{d s }\right)^{-1} \delta(\xi(s) - \eta(t))
\label{curlEtilde1}
\end{eqnarray}
Here in the delta function   $\delta(\xi(s)-\eta(t))$, 
  $\eta(t)$ can be  viewed as   a little segment, such that it  
coincides with $\xi(s)$ for $s = t_- \rightarrow t_+$.  Now  using these   segmental loop space quantities, we can express the functional divergence of the    dual loop space variable  $ \tilde{E}_\mu[\eta|t]$ as 
\begin{eqnarray}
 & & \epsilon^{\lambda\mu\alpha\beta} \delta_\lambda(t)
   \Big\{\omega^{-1}(\eta(t)) \tilde{E}_\mu[\eta|t] \omega(\eta(t))\Big\} \nonumber \\
& = &\!\!\!\!-\frac{2}{\bar{N}} \int\delta\xi ds\Big \{ \frac{d \eta^\beta}{dt } \frac{d \eta^\alpha}{d s  }  -  \frac{d \eta^\alpha}{dt }   \frac{d \eta^\beta}{d s  } \Big \} \nonumber  \\ && 
   \delta_\rho(s) \frac{\delta}{\delta \xi_\rho (s)} E^\rho[\xi|s] \left(\frac{d \xi^\gamma}{d s }  \frac{d \xi_\gamma}{d s } \right)^{-1} \delta(\xi(s)-\eta(t)).
\label{curlEtilde2}
\end{eqnarray}
multiplying this by  $  \epsilon_{\mu\nu\alpha\beta}/2$. So,   dual loop space variable, we can  write 
\begin{eqnarray}
& & \omega^{-1}(\eta(t)) \Big\{\frac{\delta }{ \delta \eta_\nu (t)} \tilde{E}_\mu[\eta|t]
   - \frac{\delta }{ \delta \eta_\mu (t)}  \tilde{E}_\nu[\eta|t]\Big\} \omega(\eta(t)) \nonumber \\
   & = & -\frac{1}{\bar{N}} \int \delta\xi ds \epsilon_{\mu\nu\alpha\beta}
  \Big \{ \frac{d \eta^\beta}{ dt } \frac{d \xi^\alpha}{ ds } - \frac{d \eta^\alpha}{ dt }
  \frac{d \xi^\beta}{ ds }\Big\} \frac{\delta}{ \delta \xi_\rho (s)} \nonumber \\ && E^\rho[\xi|s]\left(\frac{d \xi^\gamma}{d s }  \frac{d \xi_\gamma}{d s } \right)^{-1}
   \delta(\xi(s)-\eta(t)) 
\end{eqnarray}
The  loop space functional  derivatives vanish for local quantities. Thus, the action of the  loop space functional derivatives on    the transformation matrix  $\omega^{-1}(\eta(t))$ and $\omega(\eta(t))$ will vanish. The functional  divergence of the loop space variable $E^\rho[\xi|s]$ can be related  to the functional curl of the dual loop space variable $ \tilde{E}_\mu[\eta|t]$. Thus, the quantity $\mathcal{Y}_\mu = D^\nu{R} _{\mu\nu}^{ ab} \Sigma_{ab}$ in general relativity   appears as  monopole in the dual gravitational theory. Furthermore, it can be   argued that the vanishing of the functional  divergence of the loop space variable   $  E_\rho[\xi|s]$ can be related to the vanishing of functional  curl of the dual loop space variable  $ \tilde{E}_\mu[\eta|t]$. So, if the term $\mathcal{Y}_\mu = D^\nu{R} _{\mu\nu}^{ ab} \Sigma_{ab}$  vanishes in general relativity, then the dual gravitational theory will not contain any  monopoles. 

We will demonstrate that this duality is  invertible. This can be done by first observing that 
\begin{eqnarray}
 && \frac{2}{\bar{N}} \epsilon^{\alpha\beta\mu\lambda} \frac{d\zeta_\beta}{du}
   \int \delta\eta dt \omega^{-1}(\eta(t)) \tilde{E}_\mu[\eta|t]
\omega(\eta(t))
   \frac{d\eta_\lambda}{dt}\\ \nonumber  && \times \left(\frac{d\eta^\sigma}{ds}
   \frac{\eta_\sigma}{ds}\right)^{-1} \delta(\eta(t)-\zeta(u)) \nonumber
\\
& = & -\frac{4}{\bar{N}^2} \epsilon^{\alpha\beta\mu\lambda}
   \epsilon_{\mu\nu\rho\sigma} \frac{d\zeta_\beta}{du}(u) \int \delta\eta dt
  \frac{d \eta_\lambda}{dt} \frac{d\eta^\nu}{dt}\left(\frac{d \eta^\gamma}{dt} \frac{d \eta_\gamma}{dt}\right)^{-1}
   \delta(\eta(t)-\zeta(u)) \nonumber \\ && \int \delta\xi ds E^\rho[\xi|s]  \frac{d\xi^\sigma}{ds}
   \left(\frac{d\xi^\gamma}{ds} \frac{d\xi_\gamma}{ds}\right)^{-1} \delta(\xi(s)-\eta(t)).
\label{invertdual1}
\end{eqnarray}
Now we  integrate  over $\dot{\eta}(t)$,  and 
obtain a factor $\bar{N} \delta_\lambda^\nu/4$. So, we can express the   right-hand of this equation as 
\begin{equation}
\frac{2}{\bar{N}} \{\delta_\rho^\alpha \delta_\sigma^\beta - \delta_\rho^\beta
   \delta_\sigma^\alpha\} \frac{\zeta_\beta}{du} \int \delta\xi ds E^\rho[\xi|s]
  \frac{d\xi^\sigma}{ds}
   \left(\frac{d\xi^\gamma}{ds} \frac{d\xi_\gamma}{ds}\right)^{-1} \delta(\xi(s)-\eta(t)).
\label{invertdual2}
\end{equation}
Now we observe that this  is anti-symmetric in the indices $\rho$ and $\sigma$. As it is possible for  $\dot{\zeta}$ and
$\dot{\xi}$  to be parallel, can equate them 
using $\delta(\xi(s)-\zeta(u))$. Thus,  using  $E^\alpha[\zeta|u]$,   this expression can be written as 
\begin{eqnarray}
\omega(\zeta(u)) E_\alpha[\zeta|u] \omega^{-1}(\zeta(u))
  &=& \frac{2}{\bar{N}} \epsilon_{\alpha\beta\mu\lambda}
   \frac{\zeta_\beta}{du} \int \delta\eta dt \tilde{E}^\mu[\eta|t] \nonumber \\ && \times 
  \frac{d\eta_\lambda}{dt}\left(\frac{d\eta^\sigma}{dt}\frac{\eta_\sigma}{dt}\right)^{-1}\delta(\eta(t)-\zeta(u))
\label{invertdual}
\end{eqnarray}
So, it is is possible to start from dual loop space variable and obtain the loop space variable of general relativity. Thus, this duality transformation satisfies all the properties of a duality transformation.

Now we will demonstrate that   this duality reduces to the usual    Hodge duality for linearized gravity.  
However, as we have analyzed the theory using spin connection as a dynamical 
variable, we will also define the linearized gravity with the spin connection 
as the dynamical variable. The curvature tensor can now be expressed in  
terms of the  the spin connection as  \cite{dynam2, dynam0,  dynam4, dynam6,  
ynam, dynam1}
\begin{eqnarray}
 R_{\mu\nu}^{ab}   =  R_{\mu\nu (lin)}^{ab}  +  \omega_\mu^{ac} \omega^b_{\nu c}- \omega_\nu^{ac} \omega^b_{\mu c},
\end{eqnarray} 
where  $R_{\mu\nu (lin)}^{ab} = \partial_\mu \omega_\nu^{ab}
 -  \partial_\nu \omega_\mu^{ab}$ is the linearized curvature tensor. It may be noted that this model of linearized gravity has been linearized with respect to the spin connection, and so it is different from the  linearized gravity, which is linearized with respect to the metric   \cite{ge01, ge02, ge12, ge14}. However, as both these theories can be used to construct the same gauge invariant observables, both of these approach to linearized gravity are equally valid. The   Hodge duality for gravity linearized with respect to the metric has produced interesting results  \cite{du12, du14, du15, du16}. Here we will investigate the Hodge duality for linearized gravity, which has been linearized with respect to the spin connection. 
This can be seen by using  $\tilde{E}_\mu[\eta|t]$ to write 
\begin{eqnarray}
\omega^{-1}(x) \tilde{R} _{\mu\nu, (lin)}^{ ab} \Sigma_{ab}(x)(x) \omega(x) = -\frac{2}{\bar{N}}
   \epsilon_{\mu\nu\rho\sigma} \int \delta\xi ds E^\rho[\xi|s]
   \left(\frac{d\xi^\sigma}{ds}  \frac{d\xi_\alpha}{ds}\right)^{-1} \nonumber \\ \times  \delta(x - \xi(s)).
\label{reducedual1}
\end{eqnarray}
Here we will   perform the integral first, and then take  the width of 
the segment in $E_\mu[\xi|s]$ to zero. Now for the  linearized gravity 
with   a the linearized curvature  ${R}_{\mu\nu (lin)}^{ ab}$  
(linearized with respect to the spin connection), we can write 
\begin{eqnarray}
{\tilde{ R}}_{\mu\nu (lin)}^{ ab} \Sigma_{ab}(x) & = & -\frac{2}{\bar{N}} \epsilon_{\mu\nu\rho\sigma}
   \int \delta\xi ds R^{\rho\alpha ab}_{(lin)} \Sigma_{ab}(\xi(s)) \frac{d \xi_\alpha}{d s } 
   \frac{d \xi^\sigma }{d s } \nonumber \\ && \times
    \left(\frac{d\xi^\gamma}{ds}   \frac{d\xi_\gamma}{ds}\right)^{-1} \delta(x-\xi(s)) \nonumber \\
   & = & - \mbox{\small $\frac{1}{2}$} \epsilon_{\mu\nu\rho\sigma}
    R^{\rho\sigma ab}_{ (lin)} \Sigma_{ab}(x).
\label{reducedual}
\end{eqnarray}
The expression on the right-hand side is exactly the Hodge dual for the 
tensor of linearized curvature 
${R} _{\mu\nu, (lin)}^{ ab} \Sigma_{ab}(x)$. However, with non-linear terms, 
this duality cannot be expressed as  Hodge  dual. 
Thus, we have shown that the Hodge duality would only hold for 
linearized gravity, linearized with respect to spin connection.

\section{Monopole Charge} 
Now as we have analyzed general relativity as a gauge theory, with spin connection as the gauge potential, it was also possible to analyze the topological defects in this gauge theory. This was done using the loop space formalism. In fact, the loop space formalism  can be used to  obtain non-abelian monopole charge in the loop space formalism. This can be  done by constructing a loop in the loop space. As this can be done for any non-abelian gauge theory, it can also be done for the gauge theory of Lorentz symmetry.  So, we will obtain gravitational  monopoles by constructing a  loop in the  loop space of spin connection. A  loop in the loop space can be constructed using the loop space  connection $ {F}^\mu [\xi|s]$.  A loop in the loop space can be     defined using  $\Sigma$, where $\Sigma$ is defined  as \cite{pq1, dual,  dual1,  pq2}
 \begin{equation}
 \Sigma : \{ \xi^\mu (t: s): s = 0 \to 2\pi, \, \,  t = 0 \to 2\pi\},  
\end{equation} 
with $ \xi^\mu (t: 0)$ and $ \xi^\mu (0:s)$ are given by 
 \begin{eqnarray}
 \xi^\mu (t: 0) = \xi^\mu (t:2\pi), && t = 0 \to 2\pi,  \nonumber \\
 \xi^\mu (0:s) = \xi(2\pi:s), && s = 0 \to 2\pi. 
\end{eqnarray}
Now it is possible to represent closed loop $ C(t)$ at every point $t$ on  $ \xi^\mu (t: s)$ as
\begin{equation}
C(t) : \{\xi^\mu (t:s), s = 0 \to 2 \pi \}. 
\end{equation}
Now as $t$ varies,  $C(t)$ traces out a closed loop. This closed loop shrinks to a point at $t=0$ and $t = 2\pi$. Now we can  construct a  loop in this loop space, using ${F}_\mu [\xi|t,s]$ as the gauge connection as   
\begin{equation}
 \Theta (\Sigma) =    P_t \exp i \int_0^{2\pi} dt \int^{2\pi}_0 
{F}_\mu [\xi|t,s] \frac{\partial \xi^\mu [\xi|t,s ]}{\partial t}.
\end{equation}
 A parameterizes surface in the spacetime is used to define 
this loop in the loop space.  If we denote by $\Omega X$ loop space, then this
(parametrized) surface can be thought  of as an element of the double loop
space $\Omega^2 X$.
This surface  encloses a volume, and  
the gravitational monopole can  be present inside such a volume. 
So, this loop of loop space, can be used to measure a gravitational 
monopole.   

It may be noted that  ${F}_\mu [\xi|t,s]$ is a  connection in the loop 
space, 
and it is used to construct the holonomy of 
the loop in loop space. Thus,  ${F}_\mu [\xi|t,s]$  in the double loop space 
is analogous to  $\omega^{ab}_\mu \Sigma_{ab}$ in the loop space. So,  
$ {G}_{\mu\nu}[\xi ( s_1, s_2)] $ in the double loop space  is  analogous 
structure to $R^{\mu\nu ab} \Sigma_{ab}$ in the loop space. 
However,    $ {G}_{\mu\nu}[\xi ( s_1, s_2)] $ can be constructed using the  
logarithmic derivative of $\Theta (\Sigma)$. Thus, it can also be viewed as 
the connection in the loop of loop space. So, the loop space curvature can 
be viewed as a connection in the loop of loop space. Now $\Theta(\Sigma)$ 
measures the gravitational monopole charge. So, for the gravitational  
monopole charge $\zeta$  enclosed by the surface $\Sigma$, we can write 
$\Theta (\Sigma) = \zeta$ \cite{t1}. Now we can write the Euclidean version 
of $SO(3, 1)$ as $SO(4)$, and we can 
construct $\zeta$ for it. It may be noted that for three dimensional gravity, 
we can  write the Euclidean version of $SO(2, 1)$ as $SO(3)$.  Since monopole 
charges are given by elements of the fundamental groups, and fundamental 
groups of $SO(3)$ and $SO(4)$ are both isomorphic to ${\bf Z}_2$, monopoles for
the two cases are represented by the same topological obstruction.
Now for $SO(3)$ 
the monopole change is $I$ when the monopole is not present, and $-I$ when it is present. 
Now for  $s_1\neq s_2$, $ {G}_{\mu\nu}[\xi ( s_1, s_2)] $ does not enclose any volume. So, we can write  $\Theta(\Sigma) =I$, where $I$ is the group identity. Thus,  for $s_1\neq s_2$ logarithmic derivative of $\Theta$
vanishes, and so there is no monopole charge. This also occurs when  $s_1 = s_2$, but the monopole world-line $Y^\rho(\tau)$ does not 
intersect with $\xi(s)$. So, even in that case, we can write  $\Theta (\Sigma)= I$.
However, when   $s_1= s_2$, and  $\xi(s)$ intersect the monopole world-line 
$Y^\rho(\tau)$, we can write  (with  $\exp\,  i\pi \kappa = \zeta$) 
\cite{t1}, 
\begin{equation}
 {G}_{\mu\nu}[\xi ( s_1, s_2)] = \frac{-\pi}{g} \int d\tau \kappa[\xi|s]  \frac{d \xi_{ \sigma }(s)}{ds} 
  \frac{d Y_{ \rho }(\tau)}{d\tau} 
 \delta(\xi(s) - Y(\tau)) \delta(s_1-s_2).
\end{equation}
So,  the monopole change for general relativity can be explicitly constructed 
using this  loop of loop space. It may be noted that even for non-abelian 
gauge theory, such non-abelian monopole  can only be investigated using 
loop space formalism  
\cite{pq1, dual,  dual1,  pq2}. As the general relativity can be studied
as a gauge theory of Lorentz group, such a monopole change in the dual theory 
can also be studied only using the double loop space.   In fact, the 
construction of such a dual theory is only possible in loop space formalism. 
Thus, to construct a dual gravitational theory  beyond linear approximation, 
we have to analyze  general relativity using the  loop space formalism.

\section{Conclusion}

In this paper,  we have  used the spin connection to  construct Polyakov 
loop space formalism for general relativity. This was done  as general 
relativity can be viewed as a gauge theory of Lorentz group. This  Polyakov 
loop was then used   to obtain  Polyakov variable as   the connection in the 
loop space. 
Furthermore, we also have constructed the loop space curvature  using 
Polyakov variable. It was   demonstrated that this loop space curvature 
is proportional to the gravitational Bianchi identity in spacetime.  Thus, 
using  the analogy with Yang-Mills monopoles, it was   argued  that this  
loop space curvature could   be used to analyze the gravitational monopoles. 
We also argued that the loop space can be constructed using a new set of loop 
variables.  These new loop variables were  used to construct a dual  
gravitational theory beyond its linear approximation. It was observed that 
a certain quantity in the original theory produced a monopole in the 
dual theory. We also demonstrated that  such a quantity in the dual theory 
produced the monopole in the original theory.  We also  explicitly 
showed that the loop space  duality reduces to Hodge duality for 
linearized gravity.  It may be noted that  we constructed the loop space 
with spin connection as the dynamical variable, and so,  we  linearized the 
gravity with respect to the spin connection. However,  both the spin 
connection and metric can be viewed as dynamical variable, as they  
both can be  used to construct gauge invariant observables. The 
advantage of using  spin connection was that the loop space formalism 
could be constructed with the spin connection in analogy with the 
Yang-Mills theory, with spin connection as the non-abelian gauge potential. 
We have also analyzed the loop in the space of loops. We used this 
double loop space formalism to construct a gravitational monopole charge. 

It may be noted that it would be interesting to construct the loop space using metric as a dynamic variable. Then it would be possible to show that this loop space reduces to the usual Hodge duality for   linearized gravity  \cite{du12, du14, du15, du16}. This can be done by first defining the loop space variable  in such a way, that the loop space connection can be related to the curvature tensor in spacetime. This  loop space connection can be related to   a new set of loop variables.  These new loop variables can then be   used to construct a dual  gravitational theory  using metric as a dynamic variable. It would also be interesting to investigate the relation between the loop space constructed with spin connection as the dynamic variable, and the loop space constructed with metric as the dynamical variable. Furthermore, as the  Hodge duality for linearized gravity has been generalized to curved spacetime \cite{cu12, cu14}, it would be interesting to investigate such limits of  loop space formalism of gravity, with metric as a dynamical variable.  It is known that the  Hodge dual gravity  has interesting applications in M-theory \cite{mt12, mt14}, so it would be interesting to generalize it to the loop space  duality in  M-theory.  It may be noted that it  has been proposed that a six dimensional phases of  M-theory can be obtained  from  five dimensional  linearized supergravity theory \cite{mt16}.  In fact,   the   Bianchi identities have been interchanged with  a suitable quantity defined in linearized gravity  for such a system \cite{mt12, mt14}. As the interchanging of the gravitational  Bianchi identities with a quantity defined in the original gravitational theory can be studied using loop space formalism, it would be interesting to investigate such phases in M-theory using loop space formalism. The loop space formalism can also be used to analyze dual generalized geometries. This is because Hodge duality has been used to construct  the dual geometry of an $E11$ generalized   geometry \cite{mt01}  and an  $E7$ generalized geometry \cite{mt02}. So, it would  be possible   to use   the  loop space to construct dual  generalized geometries. This loop space formalism can also be used to investigate topological defects in generalized geometries. As these generalized   geometries are important in M-theory, and so the loop space formalism can be used to investigate interesting structures in M-theory.  

It would be interesting  to analyze a deformation of this  theory, by analyzing a deformation of general relativity. In fact, the deformation of a non-abelian gauge theory by a minimal length in the background geometry of spacetime has been investigated using the loop space formalism \cite{td12}.
It was demonstrated that even if the original gauge theory did not have a topological defect, it was possible for the deformed gauge theory to have a topological defect due to a minimal length in spacetime.  The  deformation of  general relativity from  such a minimal length has been investigated using spin connections \cite{dynam2}. So, it 
 would be interesting to  construct   the loop space variable for this 
deformed gravitational theory. This deformed   loop space variable could  
then be used to obtain  
a  deformed  loop space curvature.  It would be interesting to analyze 
the effect of such   deformation   on gravitational monopoles. 
It is expected that the deformation of the general relativity by a 
minimal length will also deform the gravitational Bianchi identities. 
Thus, even if the original Bianchi identity is satisfied, it could be 
violated due to such a minimal length. So, it would be possible for the 
gravitational loop space curvature not to vanish, even in absence of a 
gravitational monopole. This could mean that a minimal length can produce 
a gravitational monopole. This has already been demonstrated for Yang-Mills 
theories \cite{td12}. So, it would be instructive to see if the same phenomenon  
occurs in general relativity.


\begin{thebibliography}{99}


\bibitem{emhd}E.~P.~Verlinde,
Nucl. Phys. B  {455}, 211  (1995)
\bibitem{emhd1}C.~T.~Hsieh, Y.~Tachikawa and K.~Yonekura,
Phys. Rev. Lett. {123}, 16, 161601 (2019)



\bibitem{p1}   A.~M.~Polyakov, Nucl. Phys. B164, 171 (1980) 

\bibitem{t1} H.~M.~Chan and S.~T.~Tsou, Some Elementary Gauge Theory Concepts, World Scientific (1993)


\bibitem{Chan:1995xr}
H.~M.~Chan, J.~Faridani and S.~T.~Tsou,
 Phys. Rev. D {53}, 7293 (1996)

\bibitem{t2} H.~M.~Chan, P. Scharbach and S.~T.~Tsou, Ann. Phys. 167 454  (1986)



\bibitem{p}   H.~M.~Chan and S.~T.~Tsou, Act. Phys. Pol. B17, 259 (1986)


\bibitem{m2frac}M.~Faizal and S.~T.~Tsou, Int. J. Theor. Phys. 54, 896 (2015) 

\bibitem{td12}M.~Faizal and S.~T.~Tsou,
Nucl. Phys. B  {924}, 588-602 (2017)


\bibitem{pqaa}  M.~Faizal, Europhys. Lett. 103, 21003 (2013) 

\bibitem{pq1a}  M.~Faizal and S.~T.~Tsou, Europhys. Lett.  107, 20008 (2014)


 \bibitem{gm12}U.~Kol and M.~Porrati, Phys. Rev. D {101}, no.12, 126009 (2020)
 \bibitem{gm14} J. J. Manjarin, JHEP 0411, 017 (2004)
\bibitem{gm15} D.~Maison and S.~L.~Liebling, Phys. Rev. Lett.  {83}, 5218  (1999)
\bibitem{gm16}S.~L.~Liebling, Phys. Rev. D {61}, 024030 (2000)
\bibitem{gm17} S.~A.~Cherkis and A.~Kapustin,
Commun. Math. Phys.  203, 713 (1999)


\bibitem{dynam2}M.~Kober,
Phys. Rev. D  {82}, 085017 (2010)
\bibitem{dynam0}M.~Camenzind,
Phys. Rev. D {18}, 1068-1081 (1978)
\bibitem{dynam4} H.~R.~Pagels, Phys. Rev. D {27}, 2299 (1983)
\bibitem{dynam6}R.~J.~Mckellar,
 J. Math. Phys.  {22}, 2934  (1981)
\bibitem{dynam} S.~Das, M.~Faizal and E.~C.~Vagenas,
Int. J. Mod. Phys. D  {27}, 14, 1847002 (2018)
\bibitem{dynam1} 
S.~Das and M.~Faizal, Gen. Rel. Grav. {50}, 7, 87 (2018)
 
 

  
 
   \bibitem{dual} M. Faizal and S. T. Tsou,  Eur. Phys. J. C 75, 316  (2015) 
 \bibitem{dual1}M. Faizal and S. T. Tsou,  Found. Phys. 45, 1421 (2015)
 
 
    \bibitem{pq1}  H.   M.   Chan, J.   Faridani and S.~T.~Tsou,
  Phys.    Rev.    D      52, 6134 (1995)



    \bibitem{pq2}  H.   M.   Chan, J.   Faridani and S.~T.~Tsou, Phys.    Rev.    D      53, 7293 (1996)
    

\bibitem{char}  O. Aharony, N. Seiberg and  Y. Tachikawa, JHEP 1308,  115 (2013) 

    \bibitem{pq4}  H.   M.   Chan, J.   Bordes and S.~T.~Tsou,
  Int.    J.    Mod.    Phys.    A      14,  2173  (1999)

    \bibitem{p9}   H.~M.~Chan and S.~T.~Tsou,
  Acta Phys.    Polon.    B      28, 3027  (1997)

    \bibitem{pq0}  H.~M.~Chan and S.~T.~Tsou ,
  Acta Phys.    Polon.    B      33, 4041  (2002)

    \bibitem{q01}  H.   M.   Chan,
  Int.    J.    Mod.    Phys.    A      16, 163 (2001)

    \bibitem{d0}   H.~M.~Chan and S.~T.~Tsou,
  Acta Phys.    Polon.    B      28, 3041  (1997)

   
    \bibitem{d1}  J.   Bordes, H.   M.   Chan, J.   Faridani, J.   Pfaudler and S.~T.~Tsou,
  Phys.    Rev.    D      58, 013004 (1998)

    \bibitem{d2}  J.   Bordes, H.   M.   Chan, J.   Faridani, J.   Pfaudler and S.~T.~Tsou,
  Phys.    Rev.    D      60, 013005 (1999)

    \bibitem{no}  J.   Bordes, H.   M.   Chan, J.   Pfaudler and S.~T.~Tsou,
  Phys.    Rev.    D      58, 053003 (1998)

  
   \bibitem{cmk}  M.   Kobayashi and T.   Maskawa,
  Prog.    Theor.    Phys.          49, 652-657 (1973) 

  
  
   \bibitem{1t}  G.   't Hooft,
  Nucl.    Phys.    B      138, 1  (1978) 

    \bibitem{d}   H.~M.~Chan and S.~T.~Tsou,  Phys.    Rev.    D      57, 2507 (1998)

    \bibitem{1p}   H.~M.~Chan and S.~T.~Tsou,
  Phys.    Rev.    D      56, 3646 (1997)


\bibitem{ge01}J.~Q.~Quach,
Phys. Rev. Lett.  {114},  8, 081104 (2015)
\bibitem{ge02}P. Szekeres, Ann. Phys. 64, 599 (1971)

\bibitem{ge12}J. Ramos, M. de Montigny, and F. Khanna,Gen. Relat. Gravit. 42, 2403 (2010)
\bibitem{ge14} R. Maartens and B. A. Bassett, Class. Quantum Grav. 15, 705 (1998)
\bibitem{du12}M. Henneaux and C.  Teitelboim, Phys.Rev. D71, 024018 (2005)
\bibitem{du14}T.~L.~Curtright, Nucl. Phys. B {948}, 114784 (2019)
\bibitem{du15}X. Bekaert, N.  Boulanger and  M. Henneaux, Phys. Rev. D 67 044010 (2003)
\bibitem{du16} C.~M.~Hull, JHEP {09}, 027 (2001)
\bibitem{cu12}T. Basile, X.  Bekaert and  N. Boulanger,  Phys. Rev. D 93, 124047 (2016)
\bibitem{cu14} N. Boulanger, A.  Campoleoni and  I. Cortese, Phys.  Lett. B 782, 285 (2018) 
\bibitem{mt12}C. M. Hull and P. K. Townsend, Nucl. Phys. B 438, 109  (1995)
\bibitem{mt14}C. M. Hull, JHEP 0012, 007 (2000) 
\bibitem{mt16}C.M. Hull, Nucl.Phys. B 583, 237 (2000) 
\bibitem{mt01}P.~West,
 JHEP {02}, 018 (2012)
\bibitem{mt02}H.~Godazgar, M.~Godazgar and H.~Nicolai,
JHEP \textbf{02}, 075 (2014)
 

\end{thebibliography}
\end{document}